\documentclass[12pt,english,aps,manuscript]{article}
\usepackage[T1]{fontenc}
\usepackage[latin1]{inputenc}
\usepackage{geometry}
\usepackage{amsmath}
\usepackage{amssymb}
\usepackage[numbers]{natbib}

\makeatletter

\usepackage{color}

\makeatletter
\newcommand{\bee}{\begin{equation}}
\newcommand{\ee}{\end{equation}}
\newcommand{\beea}{\begin{eqnarray}}
\newcommand{\eea}{\end{eqnarray}}

\makeatother

\makeatother

\usepackage{babel}
\begin{document}
\begin{center}
\textbf{\Large{}Constraints on Dbar Uplifts}
\par\end{center}{\Large \par}

\begin{center}
\vspace{0.3cm}
 
\par\end{center}

\begin{center}
{\large{}S. P. deAlwis}
\par\end{center}{\large \par}

\begin{center}
Physics Department, University of Colorado, \\
 Boulder, CO 80309 USA, dealwiss@colorado.edu
\par\end{center}

\begin{center}
\vspace{0.3cm}
 
\par\end{center}

\begin{center}
\textbf{Abstract} 
\par\end{center}

\begin{center}
\vspace{0.3cm}
We discuss constraints on KKLT/KKLMMT and LVS scenarios that use anti-branes
to get an uplift to a deSitter vacuum, coming from requiring the validity
of an effective field theory description of the physics. We find these
are not always satisfied or are hard to satisfy.
\par\end{center}

Flux compactifications of type IIB string theory (for reviews see\citep{Grana:2005jc,Douglas:2006es}),
have yielded a concrete framework within which to discuss beyond the
standard model phenomenology and cosmology. The most popular scenarios
involve a first stage with an Anti-deSitter minimum for the moduli
potential \citep{Kachru:2003aw} \citep{Balasubramanian:2005zx} which
is then commonly raised to a deSitter minimum by introducing one or
more anti-D3 branes at the bottom of a warped throat in the internal
space geometry as in \citep{Kachru:2003aw} \footnote{There are of course exceptions - for a recent comprehensive review
of string cosmology see \citep{Baumann:2014nda}}. In this note we discuss the validity of certain cosmological scenarios
within this framework and estimate the one-loop quantum corrections
to soft masses, which in principle could be large due to the fact
that the SUSY breaking scale is effectively above the ultra-violet
(UV) cutoff.

The SUGRA potential in the presence of a D-bar brane (down a warped
throat) is of the form
\begin{eqnarray}
V(\tau,\Phi) & = & |F(\tau,\Phi)|^{2}-3m_{3/2}^{2}(\tau,\Phi)M_{P}^{2}+nV_{\bar{D}3}^{(w)}(\tau,\Phi)\nonumber \\
 & = & |F(\tau,\Phi))|^{2}-3m_{3/2}^{2}(\tau,\Phi)M_{P}^{2}+(M_{s}^{(w)}(\tau,\Phi))^{4}+\ldots\label{eq:Dbarpot}
\end{eqnarray}
Here $F$ stands for the F-terms of the moduli and $m_{3/2}(\tau,\Phi)$
is the moduli dependent gravitino mass and $\tau$ volume modulus.
Also $n$ is the number of D3-bar branes each with potential $V_{\bar{D}3}^{(w)}$,
the superscript $w$ indicating the warping. In the ground state of
the Dbar brane this potential is just the fourth power of $M_{s}^{(w)}$
the warped string scale\footnote{Note that the 10D equations determine the warp factor at the end of
the throat in terms of the complex structure moduli.}. Note that such a term is absent if only D branes (and orientifold
planes) are present.

This potential makes sense only below the lightest Kaluza-Klein (KK)
scale in the problem which is the scale set by KK modes localized
at the end of the warped throat. This is necessarily parametrically
smaller than the warped string scale, since all effective field theory
(EFT) discussions in string theory, depend on proceeding via 10D SUGRA
(after freezing string modes). i.e. we need to ensure,
\begin{equation}
V(\tau,\Phi)\ll(M_{KK}^{(w)}(\tau))^{4}\ll(M_{s}^{(w)}(\tau)^{4}.\label{eq:KKbound}
\end{equation}
In the literature on string inflation the only necessary criterion
that is usually imposed is that the Hubble scale is well below the
(warped) KK scale i.e. $H\ll M_{KK}^{(w)}$. However the EFT must
describe not only the inflationary regime but also the dynamics at
the end of inflation - namely the reheating phase. It is easy to see
though that if the scale of the inflationary potential is higher than
the (warped) KK scale then this will lead to kinetic energies of the
inflaton(s) that is higher than the (warped) KK scale, thus violating
the criterion under which the derivative expansion is valid\footnote{See for example section 2.1 and footnote 6 of \citep{Burgess:2016owb}.}. 

One can certainly ensure this at the minimum of the potential - indeed
that is the role of the last (uplift) term (which is greater than
the warped KK scale) 
\[
V_{min}=V(\tau_{0},\Phi_{0})=|F(\tau_{0},\Phi_{0}))|^{2}-3m_{3/2}^{2}(\tau_{0},\Phi_{0})M_{P}^{2}+(M_{s}^{(w)}(\tau_{0},\Phi_{0}))^{4}\gtrsim0.
\]
 However it is not possible to ensure this away from the minimum without
functional fine-tuning. As one moves away from the minimum it becomes
harder and harder to satisfy the bound \eqref{eq:KKbound} since the
scale of the potential is set by the warped string scale which should
be much larger than the warped KK scale (see below for estimates of
these). 

From the above it appears that there is no consistent EFT picture
of inflation in the presence of D-bar uplifts, when the inflaton is
one of the fields in the EFT. The LVS case with so-called ``Fibre
inflation'' is however an exception and will be discussed later.

On the other hand inflationary dynamics in constructions (such as
KKLT, KKLMMT \footnote{For a recent review with a comprehensive discussion of all the issues
involving this type of scenario see Baumann and McAllister \citep{Baumann:2014nda}.}) typically involve the motion of branes moving down the warped throat
with inflation ending and a reheating phase taking over once the brane
and anti-brane annihilate. Such a process will end with the creation
of (light warped) string states and KK states with masses respectively
$M_{s}^{(w)}$ and $M_{KK}^{(w)}$. But this is clearly a stringy
process taking place at energy scales $E\sim M_{s}^{(w)}\gg M_{KK}^{(w)}$.
Clearly one cannot discuss this within a 4D low energy effective field
theory. In this case the above potential is just used to establish
the background field configuration in which this intrinsically stringy
process takes place. 

To incorporate warping we adopt the following parametrization of the
10D string metric. This is related to the 4D Einstein metric $g_{\mu\nu}$
by\footnote{Note that the standard factorized metric used in GKP cannot be used
everywhere on the ten dimensional space. In addition in the presence
of warping one needs to keep non diagonal terms $G_{\mu n}$ in the
metric ansatz - since otherwise the Einstein equation $R_{\mu n}=0$
leads to an overconstrained system as first observed in \citep{deAlwis:2003sn}
- see also \citep{deAlwis:2004qh,Giddings:2005ff}.}
\begin{eqnarray}
ds^{2} & = & G_{MN}dX^{M}dX^{N}\nonumber \\
 & = & e^{\phi/2}[e^{-6u(x)}\left(1+\frac{e^{-4A(y)}}{e^{4u(x)}}\right)^{-1/2}g_{\mu\nu}(x)dx^{\mu}dx^{\nu}+\nonumber \\
 &  & e^{2u(x)}\left(1+\frac{e^{-4A(y)}}{e^{4u(x)}}\right)^{1/6}\hat{g}_{mn}(y)dy^{m}dy^{n}]\label{eq:metric}
\end{eqnarray}
Here $\phi$ is the dilaton, $g_{mn}$ is a fiducial metric on the
internal manifold $X$ which is conveniently normalized such that
$\int_{X}\sqrt{\hat{g}}d^{6}y=(2\pi)^{6}\alpha'^{3}$. So we have
(using the relation $2\kappa_{10}^{2}=(2\pi)^{7}\alpha'^{4}$ between
the 10 D gravitational constant and the string scale in the string
metric \citep{deAlwis:1996ez})
\begin{equation}
\frac{1}{(2\pi)^{7}\alpha'^{4}}\int d^{10}X\sqrt{|G|}e^{-2\phi}G^{\mu\nu}R_{\mu\nu}(G)+\ldots=\frac{1}{2\pi\alpha'}\int d^{4}x\sqrt{|g(x)|}g^{\mu\nu}R_{\mu\nu}(g)+\ldots.\label{eq:4D}
\end{equation}
The Planck length $l_{P}$ measured in the Einstein frame is given
by $l_{P}^{2}=M_{P}^{-2}=2\pi\alpha'/2$. Outside of throat regions
the warping can be ignored so that $e^{4u}\gg e^{-4A}$ and the metric
becomes the standard 4D Einstein metric as given for example in GKP\citep{Giddings:2001yu}.
The volume modulus is then $\tau=e^{4u(x)}$ so that the dimensionless
volume of $X$ is given by ${\cal V}=e^{6u}$. Even in the case where
there is one (or more) warped throat, as long as the volume of warped
regions is negligible compared to the overall volume, we may still
take ${\cal V}=e^{6u(x)}\int d^{6}y\left(1+\frac{e^{-4A(y)}}{e^{4y(x)}}\right)^{1/6}/(2\pi)^{6}\alpha'^{3}\simeq\tau^{3/2}=e^{6u}$.
However at the bottom of a long warped throat $y\rightarrow y_{0})$
where $e^{-4A(y_{0})}\gg e^{4u}$ the metric \eqref{eq:metric} becomes
$ds^{2}=e^{\phi/2}[e^{-4u}e^{2A}g_{\mu\nu}dx^{\mu}dx^{\nu}+e^{4u/3}e^{-2A(y)/3}\hat{g}_{mn}(y)dy^{m}dy^{n}]$.
The effective KK radius depends on the location in the internal space.
Thus this parametrization provides an interpolation between the unwarped
bulk region and the warped throat region as discussed in \citep{Kachru:2003sx}.

The string scale for strings located at a fixed value of the internal
coordinate $y$ can be read off from the string action which reads
(using \eqref{eq:metric}),
\begin{eqnarray*}
\frac{1}{2}\frac{1}{2\pi\alpha'}\int_{{\rm ws}}d^{2}\sigma\sqrt{\det(G_{MN}\partial_{\alpha}X^{M}\partial_{\beta}X^{N})} & = & \frac{1}{2}\frac{1}{2\pi\alpha'}\int_{{\rm ws}}d^{2}\sigma e^{\phi(x(\sigma),y)/2}e^{-6u(x(\sigma)}\left(1+\frac{e^{-4A(y)}}{e^{4u(x(\sigma))}}\right)^{-1/2}\\
 &  & \times\sqrt{\det(g_{\mu\nu}\partial_{\alpha}x^{\mu}(\sigma)\partial_{\beta}x^{\nu}(\sigma))},
\end{eqnarray*}
 so that relative to the 4D Einstein metric the string scale is (with
$<>_{{\rm ws}}$ denoting average over the world sheet),
\begin{equation}
M_{s}^{2}=\frac{1}{2\pi\alpha'}<e^{\phi/2}e^{-6u(x)}\left(1+\frac{e^{-4A(y)}}{e^{4u(x(\sigma))}}\right)^{-1/2}>{}_{{\rm ws}}.\label{eq:stringscale1}
\end{equation}
 Thus we have in the different regions:
\begin{eqnarray}
M_{s}^{2} & = & \frac{1}{2\pi\alpha'}e^{\phi/2}e^{-6u(x)}=M_{P}^{2}\frac{e^{\phi/2}}{2{\cal V}},\,\,{\rm unwarped};\label{eq:Msunwarped}\\
M_{s}^{({\rm W})2} & = & \frac{e^{\phi/2}}{2\pi\alpha'}{\rm e^{-4u(x)}e^{2A(y)}=M_{P}^{2}\frac{e^{\phi/2}}{2{\cal V}^{2/3}}e^{2A(y)}.}\,\,{\rm warped\,KKLMMT}\label{eq:Mswarped}
\end{eqnarray}

The KK mass matrix turns out to be, after expanding $\Phi(x,y)=\sum\phi_{r}(x)\omega_{s}(y)$
where $\omega_{r}(y)$ is an orthonormal (in the metric $\hat{g}$)
basis of harmonics on $X$, 
\[
M_{KKrs}^{2}=\frac{1}{(2\pi)^{7}\alpha'^{4}}\int_{X}d^{6}y\sqrt{\hat{g}}e^{-8u(x)}\left(1+\frac{e^{-4A(y)}}{e^{4u(x)}}\right)^{-2/3}\hat{g}^{mn}(y)\partial_{m}\omega_{r}(y)\partial_{n}\omega_{s}(y).
\]
Thus we have in the different regions, ignoring an $O(1)$ matrix;
\begin{eqnarray}
M_{KK}^{2} & = & \frac{1}{2\pi\alpha'}e^{-8u(x)}=\frac{M_{P}^{2}}{2{\cal V}^{4/3}},\,{\rm unwarped}\label{eq:MKKunwarped}\\
M_{KK}^{({\rm W})2} & = & \frac{1}{2\pi\alpha'}e^{-16u(x)/3}e^{8A(y_{0})/3}=\frac{M_{P}^{2}e^{8A(y_{0})/3}}{2{\cal V}^{8/9}},\,\,{\rm warped\,KKLMMT}.\label{eq:MKKwarped}
\end{eqnarray}
Note that the last two relations applies to modes $\omega_{r}(y)$
that have support in a strongly warped throat $y\sim y_{0}$. We observe
also that 
\[
\frac{M_{KK}^{2}}{M_{s}^{2}}=\frac{e^{-\phi/2}}{{\cal V}^{1/3}};\,\frac{M_{KK}^{({\rm W})2}}{M_{s}^{({\rm W})2}}=\frac{e^{-\phi/2}e^{2A(y_{0})/3}}{{\cal V}^{2/9}}.
\]
The last two relations apply to the unwarped and warped (KKLMMT) cases
respectively and clearly both satisfy the necessary criterion that
the KK scale should be well below the string scale for large volume
and warping. The requirement that the gravitino mass is less than
the (warped) KK then gives (using $m_{3/2}^{2}=g|W_{0}|^{2}M_{P}^{2}/{\cal V}^{2}$
and eqn \eqref{eq:MKKwarped}) 
\[
\frac{g^{3/2}|W_{0}|^{3}}{{\cal V}^{5/3}}\ll e^{4A(y_{0})}\ll\frac{1}{{\cal V}^{2/3}},
\]
 where the second relation is the statement of being in the region
of large warping (see discussion below eqn \eqref{eq:4D}.

For the corrected (KKLMMT) KKLT stabilization mechanism where the
supersymmetric AdS minimum is uplifted to Minkowski/dS space, the
relevant uplift condition gives 
\[
3M_{P}^{4}\frac{e^{\phi}|W_{0}|^{2}}{{\cal V}^{2}}\simeq M_{s}^{(W)4}=M_{P}^{4}\frac{e^{\phi}}{4{\cal V}^{4/3}}e^{4A(y_{0})},
\]
 leading to
\begin{equation}
e^{4A(y_{0})}=12\frac{W_{0}^{2}}{{\cal V}_{0}^{2/3}}\ll1.\label{eq:KKLTuplift}
\end{equation}

Let us now discuss whether adding an anti-brane contribution to the
potential can give us Minkowski (or dS) space when used in conjunction
with the spontaneously broken supersymmetric, albeit AdS, equilibrium
(LVS) solution of BBCQ\citep{Balasubramanian:2005zx}. If the scale
of the inflationary potential is given by the warped string scale,
as would be the case generically for Dbar uplift theories (see the
discussion above eqn \eqref{eq:KKbound}), then the same conclusion
applies and there is no 4D EFT that covers inflation and reheating.
However unlike in the KKLT case in LVS there is another mechanism
for inflation - namely the so-called ``Fibre Inflation'' model.
In this case (for details see \citep{Cicoli:2008gp,Burgess:2016owb}
) one starts with CY manifolds that are (for instance) K3 fibre bundles.
Compared to the original BBCQ class of models there is in addition
to the volume modulus, (at least) one more large modulus in addition
to the small (exceptional divisor) modulus $\tau_{s}$. The canonical
example is ${\cal V}=a\sqrt{\tau^{1}}\tau^{2}-b\tau_{s}^{3/2}.$ 

In this case there is at least one flat direction that survives the
BBCQ (plus uplift to dS space) stabilization procedure, that can be
used to get an inflationary scenario. This is stabilized by adding
in string loop and/or $\alpha'^{3}$ terms that result in $F^{4}$
terms. This gives a viable EFT description of inflation satisfying
all the observational constraints though predicting very low power
in tensor modes. For us the relevant question is whether this mechanism
survives the use of anti-branes to accomplish the uplift to dS.

The new element here is that there is a relation in the LVS construction
amongst the three terms of \eqref{eq:Dbarpot} coming from the extremization
with respect to the small modulus $\tau_{s}$ and the volume ${\cal V}$.
The inflationary potential then has essentially a fixed value of ${\cal V}$
but the important point is that there are cancellations amongst the
three terms, so that the scale of inflation is now given by,
\[
V_{{\rm inf}}\simeq\frac{gW_{0}^{2}\Phi(\ln{\cal V}/W_{0})}{{\cal V}^{3}}M_{P}^{4}\ll M_{KK}^{(W)4}=\frac{e^{16A(y_{0})/3}}{4{\cal V}^{16/9}}M_{P}^{4},
\]
where $\Phi\sim(\ln({\cal V}/W_{0}))^{3/2}$ and in the second relation
we have again imposed the bound \eqref{eq:KKbound}. This gives us
\begin{equation}
e^{4A_{0}}\gg\frac{(gW_{0}^{2}\Phi)^{3/4}}{{\cal V}^{11/12}}.\label{eq:A0bound}
\end{equation}
On the other hand the LVS (plus Dbar uplift term) stabilization conditions
give\footnote{Up to $O(1)$ factors the LVS potential with Dbar uplift term after
fixing the complex structure and the dilaton, takes the form, $\frac{4}{3}g(a|A|)^{2}\frac{\sqrt{\tau_{s}}e^{-2a\tau_{s}}}{{\cal V}}-2ga|AW_{0}|\frac{\tau_{s}e^{-a\tau_{s}}}{{\cal V}^{2}}+\frac{3}{8}\frac{\xi|W_{0}|^{2}}{g^{1/2}{\cal V}^{3}}+\frac{g}{2{\cal V}^{4/3}}e^{4A(y_{0})}$.
The equations \eqref{eq:tau_s} come from the extremization conditions
$\partial_{\tau^{s}}V=\partial_{{\cal V}}V=0$.} 
\begin{equation}
\tau_{s}^{3/2}\simeq\frac{\xi}{2\sqrt{g^{3}}}+\frac{16}{27}\frac{e^{4A_{0}}}{|W|^{2}}{\cal V}^{5/3}\simeq\left(a^{-1}\ln\frac{{\cal V}}{|W_{0}|}\right)^{3/2}.\label{eq:tau_s}
\end{equation}
Combining this with the bound \eqref{eq:A0bound} we get a bound on
the volume, 
\begin{equation}
{\cal V}\ll\left(\frac{27}{16}\right)^{4/3}\frac{|W_{0}|^{2/3}}{g\Phi(\ln{\cal V}/W_{0})}(a^{-1}\ln{\cal V}/W_{0}))^{2}\sim\frac{|W_{0}|^{2/3}}{ga^{2}}(\ln\frac{{\cal V}}{W_{0}})^{1/2}.\label{eq:volbound}
\end{equation}
Since typically $W_{0}\lesssim O(1),\,g\sim10^{-1}$ in LVS constructions
this bound is hard to satisfy for the typically large values ${\cal V}\gtrsim10^{3}$
that are usually required in LVS constructions.  

Recently backgrounds involving an anti-brane have been discussed in
terms of non-linearly realized SUSY (see for example \citep{Kallosh:2015nia}
and references therein), so that its effect is represented by a Goldstino
field. The latter in turn comes from a nilpotent superfield $X$ which
satisfies $X^{2}=0$. Adding the contribution of this field to the
usual Kaehler and superpotential terms we have\footnote{This is a somewhat modified version of the discussion in \citep{Kallosh:2015nia}.}
(with $k(U,\bar{U})\equiv i\int_{X}\Omega\wedge\bar{\Omega}$) and
$M_{P}=1$)
\begin{eqnarray}
K & = & -2\ln{\cal V}-\ln(S+\bar{S})-\ln k(U,\bar{U})+\frac{c(U,\bar{U},S_{R})}{{\cal V}^{\beta}}X\bar{X}+\ldots,\label{eq:K}\\
W & = & M^{2}X+W_{{\rm flux}}+W_{{\rm np}}+W_{{\rm sm}}.\label{eq:W}
\end{eqnarray}
The last term $W_{{\rm sm}}$ is the supersymmetric standard model
superpotential. The additional term coming from the anti-brane is
then\footnote{Note that we are taking dimensionless moduli so $K,W,M,X$ are all
dimensionless.}, $M_{P}^{-4}|F^{X}|^{2}=e^{K}K^{X\bar{X}}M^{4}=\frac{e^{\phi}c^{-1}{\cal V}^{\beta}}{2{\cal V}^{2}k(U,\bar{U})}M^{4}.$
Comparing $|F^{X}|^{2}$ with the leading term of the $\bar{D}_{3}$
brane action in the Einstein frame $M_{s}^{w)4}$ (see \eqref{eq:Mswarped})
and using the fact that $M$ has to be holomorphic or a constant (since
it is a term in the superpotential) we find in the KKLMMT case, $M^{4}=\frac{1}{2\pi}e^{4A_{0}},\,\beta=2/3,\,c=\frac{e^{\phi}}{k(U,\bar{U})}$.
With these identifications the low energy effects of the anti-brane
can clearly be represented within the formal context of ${\cal N}=1$
SUGRA\footnote{For an earlier discussion using spurion instead of nil-potent superfields
see \citep{Choi:2005ge}.}. 

However a formal representation does not imply that the usual properties
of linearly realized and spontaneously broken SUSY hold here. Indeed
SUSY is broken above the cutoff scale of the EFT. $F_{X}\simeq(M_{s}^{(w)})^{2}>(M_{KK}^{(w)})^{2}\equiv\Lambda_{{\rm cutoff}}^{2}$.
Thus perturbative corrections to masses couplings etc will not have
SUSY cancellations in the EFT and will be no different from that in
a non-supersymmetric theory cutoff at $M_{KK}^{(w)}$. In particular
one might expect that soft scalar masses (and hence the Higgs mass
in particular), will acquire quantum corrections $\Delta m_{0}^{2}\simeq\Lambda_{{\rm cutoff}}^{2}/16\pi^{2}$
and that the classical calculation has no meaningful phenomenological
consequences.

The actual situation is however is somewhat more complicated. It will
turn out that although the phenomenology coming from a KKLT type stabilization
(with the flux superpotential taking extremely tiny values ($W_{0}\sim e^{-{\cal V}^{2/3}}$)
will acquire one-loop corrections that tend to vitiate the classical
calculations, in the LVS case (with $W_{0}\sim O(1)$) these corrections
are actually suppressed.

From the Coleman-Weinberg formula the largest supersymmetric standard
model field dependent contribution to the effective potential is 
\begin{equation}
\Delta V_{1}=\frac{\Lambda^{2}}{16\pi^{2}}{\rm Str}{\cal M}^{2}(\Phi,\bar{\Phi}),\label{eq:CWpot}
\end{equation}
where ${\rm Str}{\cal M}^{2}=\sum(-1)^{2j+1}(2j+1){\rm tr}m_{j}^{2}(\Phi,\bar{\Phi})$.
Here $m_{j}^{2}$ is the mass matrix for states of spin $j$ and $\Phi=\{\Phi^{I}\},\,I=1,\ldots,N_{Tot}$)
stands for all the fields in the low energy theory below the cutoff
$\Lambda$, i.e. in our case all the moduli and the dilaton as well
as the matter fields. 

Now in a (spontaneously broken) supersymmetric theory (with all states
forming complete supermultiplets below the cutoff) we have \citep{Wess:1992cp}\citep{Ferrara:1994kg},
\[
{\rm Str}{\cal M}^{2}=(N_{{\rm Tot}}-1)m_{3/2}^{2}(\Phi,\bar{\Phi})-F^{I}({\cal R}_{I\bar{J}}+S_{I\bar{J}})F^{\bar{J}},
\]
where ${\cal R}_{I\bar{J}}=\partial_{I}\partial_{J}\ln\det K_{M\bar{N}}$
and $S_{I\bar{J}}=-\partial_{I}\partial_{\bar{J}}\ln\det\Re f_{ab}$
with $f_{ab}$ being the gauge coupling superfield. The field dependent
gravitino mass is given by $m_{3/2}^{2}(\Phi,\bar{\Phi})=e^{K}|W(\Phi)|^{2},$
where $K$ and $W$ are given by \eqref{eq:K}\eqref{eq:W}. For future
reference we also write the matter superpotential as 
\begin{equation}
W_{{\rm sm}}=\mu_{ij}(U)C^{i}C^{j}+\lambda_{ijk}(U)C^{i}C^{j}C^{k}+O(C^{4}),\label{eq:Wmatter}
\end{equation}
where $C^{i}$ represent the Higgs, lepton and quark superfields.

In this case the correction to the matter field mass matrix is (see
for example \citep{Choi:1997de}) 
\[
\Delta m_{i\bar{j}}^{2}=-\frac{\Lambda^{2}}{16\pi^{2}}\left(e^{K}D_{i}D_{M}W{\cal R}^{M\bar{L}}D_{\bar{j}}D_{\bar{L}}\overline{W}+O((m_{3/2}^{2},F)\right).
\]
 The only contribution in the above expression that is not proportional
to supersymmetry breaking parameters or to the matter fields $C^{i}$
(which will have vanishingly small expectation values) is the $\mu$-term
contribution from \eqref{eq:Wmatter}. In other words we have (restoring
$M_{P}$ and the dimensionality of $\mu$), 
\begin{equation}
\Delta m_{i\bar{j}}^{2}=-\frac{\Lambda^{2}}{16\pi^{2}M_{P}^{2}}\left(e^{K}\mu_{ik}\bar{\mu}_{\bar{j}l}{\cal R}^{k\bar{l}}+O(C,m_{3/2}^{2},F)\right).\label{eq:muterm}
\end{equation}
Since the $\mu$-term needs to be tuned in any case to be of the same
order as the soft SUSY breaking masses, the above quantum correction
to the squared soft mss is negligible for cut-offs $\Lambda$ which
are well below the Planck scale - as is of course the case here.

On the other hand when as here, the supersymmetry breaking is above
the cutoff, one might on general grounds expect a much larger quantum
(additive) correction to the squared soft mass. The point is that
in this case we do not have a complete supermultiplet for the goldstino
field $X$. The scalar superpartner is absent. It corresponds to the
(would be) modulus field corresponding to the position of the $\bar{D}3$
brane at the bottom of the throat and which has been removed by the
orientifold projection. Thus the supertrace Coleman-Weinberg formula
\eqref{eq:CWpot} will contain an unpartnered contribution ${\rm tr}m_{X\bar{X}}^{2}$
giving us an additional contribution to \eqref{eq:CWpot} \footnote{Note that this is a field dependent contribution to the potential
which needs to be differentiated with respect to $C,\bar{C}$ to get
the corresponding correction to soft terms. Of course on-shell (i.e.
at the minimum of the potential) this will vanish as it is the sqared
Goldstino mass.} 
\[
\Delta_{X}V_{1}=\frac{\Lambda^{2}}{16\pi^{2}}{\rm tr}m_{X\bar{X}}^{2}=\frac{\Lambda^{2}}{16\pi^{2}}e^{K}D_{X}D_{I}WD_{\bar{X}}D_{\bar{J}}\bar{W}K^{I\bar{J}}K^{X\bar{X}}.
\]
For simplicity we will specialize to the case of one Kaehler modulus
$T$ so that ${\cal V}\sim\tau^{3/2},\,\tau=\Re T$, and also ignore
the indices on the matter fields $C^{i}$. Thus we rewrite the Kaehler
potential \eqref{eq:K} as 
\begin{eqnarray*}
K & = & -3\ln(T+\bar{T})-\ln(S+\bar{S})-\ln k(U,\bar{U})+\\
 &  & \frac{aC\bar{C}}{T+\bar{T}}+\frac{cX\bar{X}}{(T+\bar{T})}+\frac{bC\bar{C}X\bar{X}}{(T+\bar{T})^{2}}+\ldots
\end{eqnarray*}
Here $a,b,c$ are functions of $U,\bar{U},\,{\rm and}\,S$. After
some straightforward calculation we find the leading contribution
proportional to $C\bar{C}$ to be\footnote{At the minimum of the potential where we expect $C=0$ the contribution
to ${\rm tr}m_{X\bar{X}}^{2}$ below will of course vanish!} 
\begin{equation}
{\rm tr}m_{X\bar{X}}^{2}\sim\frac{1}{{\cal V}^{2}}\left\vert \frac{\partial\mu^{2}}{\partial U}\right\vert ^{2}K^{U\bar{U}}K^{X\bar{X}}\sim\frac{1}{{\cal V}^{2}}\left\vert \frac{\partial\mu^{2}}{\partial U}\right\vert ^{2}K^{U\bar{U}}\frac{b}{c^{2}}C\bar{C}.\label{eq:trmXXbar}
\end{equation}
Hence the largest one-loop contribution to the squared soft mass is
\[
\Delta m_{C\bar{C}}^{2}\sim\frac{\Lambda^{2}}{32\pi^{2}}\frac{1}{{\cal V}^{2}}\left\vert \frac{\partial\mu^{2}}{\partial U}\right\vert ^{2}K^{U\bar{U}}\frac{b}{c^{2}}.
\]
 The $\mu$-term at the minimum of the potential (i.e. with $U=U_{0}$)
needs to be tuned to be of the order of the weak mass scale (or at
most $O(m_{3/2})$) so that standard model particles are at the right
scale, but in general away from the minimum it is $O(1)$ on the Planck
scale. Thus identifying the cutoff with the KK-scale (given in \eqref{eq:MKKwarped})
we have
\begin{equation}
\Delta m_{C\bar{C}}^{2}\sim\frac{M_{P}^{2}}{32\pi^{2}}\frac{e^{8A/3}}{{\cal V}^{8/9}}\frac{O(1)}{{\cal V}^{2}}\sim\frac{O(1)}{32\pi^{2}}\frac{m_{3/2}^{2}}{W^{2/3}}\frac{1}{{\cal V}^{4/3}}.\label{eq:Deltam2KKLT}
\end{equation}
In the last relation we have used the KKLT uplift condition \eqref{eq:KKLTuplift}
to estimate the warp factor. The classical squared soft mass in this
class of models is $m_{C\bar{C}}^{2}\sim m_{3/2}^{2}/|\ln m_{3/2}|$
and requiring that the quantum contribution \eqref{eq:Deltam2KKLT}
does not dominate this gives us the condition 
\begin{equation}
|\ln m_{3/2}|<32\pi^{2}W_{0}^{2/3}{\cal V}^{4/3}.\label{eq:lnm3/2}
\end{equation}
The Dbar brane contribution uplifts the first stage SUSY AdS minimum
of KKLT to Minkowski/dS space. This procedure however works only with
highly suppressed values of the flux superpotential - indeed the first
stage results in $D_{T}W=0$ which in turn gives (writing $W_{{\rm np}}=De^{-aT}$)
\begin{equation}
-aDe^{-aT}=3\frac{W_{0}}{T+\bar{T}}.\label{eq:KKLTstab}
\end{equation}
Thus we get $|\ln m_{3/2}|\sim\left\vert \ln\frac{W_{0}}{{\cal V}}\right\vert \sim a\tau.$
Using this and \eqref{eq:KKLTstab} in the bound \eqref{eq:lnm3/2}
we have $e^{2a\tau/3}<32\pi^{2}\tau^{5/3}$ . This restricts the value
of the modulus to the range $1\ll a\tau\sim\tau\lesssim15$.

In the LVS case however there is no such restriction since the uplift
condition is different. Before the uplift the potential $\sim-e^{\phi}|W_{0}|^{2}/{\cal V}^{3}$.
Requiring that this is of the same order as the warped string scale
raised to the fourth power we have instead of \eqref{eq:KKLTuplift}
the estimate $e^{4A_{0}}=\frac{|W_{0}|^{2}}{{\cal V}^{5/3}}.$ Using
this in the first relation of\eqref{eq:Deltam2KKLT} we get $\Delta m_{C\bar{C}}^{2}=\frac{O(1)}{32\pi^{2}}\frac{m_{3/2}^{2}}{W_{0}^{2/3}}\frac{1}{{\cal V}^{2}}.$ 

In the unsequestered LVS case \citep{Conlon:2005ki}, one gets soft
masses of the same order as in the KKLT case. Requiring again that
the quantum correction does not dominate the classical one now gives,
\[
|\ln m_{3/2}|\sim\left\vert \ln\frac{W_{0}}{{\cal V}}\right\vert <32\pi^{2}|W_{0}|^{2/3}{\cal V}^{2},
\]
which is easily satisfied since unlike in the KKLT case $W_{0}$ is
not required to be exponentially suppressed in LVS. In the sequestered
scenario \citep{Blumenhagen:2009gk,deAlwis:2009fn,Aparicio:2014wxa}
the classical squared soft mass is suppressed by an extra factor of
${\cal V}$ so that the above relation is replaced by 
\[
|\ln m_{3/2}|\sim\left\vert \ln\frac{W_{0}}{{\cal V}}\right\vert <32\pi^{2}|W_{0}|^{2/3}{\cal V}.
\]
 This is also easily satisfied.

In conclusion we find that with Dbar uplifts the constraints coming
from the validity of an EFT description depend on whether one is in
the KKLT case or the LVS case. In the first (KKLT) case they indicate
there is no EFT which describes inflationary cosmology without functional
fine-tuning. It is necessarily a stringy process coming from brane
anti-brane annihilation with the EFT just determining the background.
Also in order for the classical phenomenology to survive one-loop
quantum effects, the Kaehler modulus needs to be less than or equal
to a number around 15. In the second (LVS) case the constraints are
somewhat different. On the one hand it appears that it is difficult
to get a viable EFT description of inflation (such as fibre inflation)
with Dbar uplifts because it is hard to satisfy \eqref{eq:volbound}.
However the classical evaluation of soft masses will not be vitiated
by large one-loop corrections even for very large values of the Kaehler
moduli.

\section*{Acknowledgements}

I wish to thank Joe Polchinski and Fernando Quevedo for discussions
and an anonymous referee for several useful comments. I also wish
to thank the Abdus Salam ICTP for hospitality during the initial stages
of this project.

\bibliographystyle{apsrev}
\bibliography{myrefs}

\end{document}